\documentclass[pra,twocolumn,floatfix,superscriptaddress,longbibliography,nofootinbib]{revtex4-2}

\usepackage{enumerate}
\usepackage[T1]{fontenc}
\usepackage[colorlinks=true,citecolor=blue,linkcolor=magenta]{hyperref}
\usepackage{microtype}
\usepackage{graphicx}
\usepackage{booktabs} 
\usepackage{makecell}
\usepackage{tabularx,tabulary}
\usepackage{underscore}

\usepackage[export]{adjustbox}

\usepackage[caption = false]{subfig}

\usepackage{amsmath}
\usepackage{amssymb}
\usepackage{mathtools}
\usepackage{amsthm}
\usepackage{dsfont}

\usepackage{physics}
\usepackage{bm}
\usepackage{bbm}

\newsavebox\CBox
\def\textBF#1{\sbox\CBox{#1}\resizebox{\wd\CBox}{\ht\CBox}{\textbf{#1}}}

\usepackage{aeguill}
\newcommand{\guillet}[1]{
	\guillemotleft {#1}\guillemotright
}

\usepackage[capitalize,noabbrev]{cleveref}

\theoremstyle{plain}

\theoremstyle{definition}

\theoremstyle{remark}

\usepackage[textsize=tiny]{todonotes}

\begin{document}

\preprint{APS/123-QED}

\title{Application of quantum-inspired generative models to small molecular datasets}

\author{Charles Moussa}
\email{c.moussa@liacs.leidenuniv.nl}
\affiliation{%
 LIACS, Leiden University, Niels Bohrweg 1, 2333 CA Leiden, Netherlands}
\author{Hao Wang}
\email{h.wang@liacs.leidenuniv.nl}
\affiliation{%
 LIACS, Leiden University, Niels Bohrweg 1, 2333 CA Leiden, Netherlands}
\author{Mauricio Araya-Polo}
\affiliation{TotalEnergies EP Research \& Technology USA}
\author{Thomas B\"{a}ck}
\affiliation{%
 LIACS, Leiden University, Niels Bohrweg 1, 2333 CA Leiden, Netherlands}%
\author{Vedran Dunjko}
\email{v.dunjko@liacs.leidenuniv.nl}
\affiliation{%
 LIACS, Leiden University, Niels Bohrweg 1, 2333 CA Leiden, Netherlands}%

\date{\today}

\begin{abstract}
Quantum and quantum-inspired machine learning has emerged as a promising and challenging research field due to the increased popularity of quantum computing, especially with near-term devices. Theoretical contributions point toward generative modeling as a promising direction to realize the first examples of real-world quantum advantages from these technologies. A few empirical studies also demonstrate such potential, especially when considering quantum-inspired models based on tensor networks. In this work, we apply tensor-network-based generative models to the problem of molecular discovery. In our approach, we utilize two small molecular datasets: a subset of $4989$ molecules from the QM9 dataset and a small in-house dataset of $516$ validated antioxidants from TotalEnergies. We compare several tensor network models against a generative adversarial network using different sample-based metrics, which reflect their learning performances on each task, and multiobjective performances using $3$ relevant molecular metrics per task. We also combined the output of the models and demonstrate empirically that such a combination can be beneficial, advocating for the unification of classical and quantum(-inspired) generative learning.

\end{abstract}

\maketitle

\section{Introduction}
\label{intro}
With quantum computing gaining popularity, quantum algorithms for Machine Learning (ML) are being developed as new methods with potential advantages over their classical counterparts. In theory, some algorithms provide computational advantages in functional and decision problems~\cite{qrecom, Allcock2020, qsvm, qeml, qmeans} as well as advantages in sampling problems~\cite{qaoasupremacy, bornsupremacy}. Yet hardware is limited in many regards, such as size (qubit numbers), gate fidelities, architecture (qubit connectivity), and qubit lifetimes (coherence times). Such limitations can prevent successful deployments of quantum algorithms on hardware. For instance, the required number of qubits can be too small for tackling an application of interest.
\par
In the early stages of exploring the capacities of quantum ML algorithms, we generally rely on simulations on a classical computer to retrieve their performances under perfect conditions~\cite{xaviperf, moussa2022resource, moussa2022unsupervised, moussa2021tabu}. This can serve as a baseline not only to analyze models for a current task but also to study the effect of running them later under different conditions~\cite{moussa2022hyperparameter, moussa2022resource, patel2022reinforcement}. For instance, we can change a hyperparameter or run models on real hardware and either improve performances or introduce new techniques to get as close as possible to the results from perfect simulations. Also, currently running quantum algorithms on real hardware can be very costly. Having baselines from classical simulations can then help in saving quantum computing costs, and even understanding beforehand what task characteristics may be more interesting and suitable for running on quantum hardware~\cite{Moussa2020}.
\par 
For generative modeling tasks, quantum-inspired models such as tensor networks (TN) can be considered before running hybrid or full quantum models, especially for problems where the required number of qubits may be high. TN-based simulations of quantum computations have also been used as generative models~\cite{unsupervisedtn}. They can also tackle large-dimensional problems (for current quantum hardware) such as molecular string-based generation. Finally, they have also been compared to classical state-of-the-art generative models, such as generative adversarial network (GAN) for combinatorial optimization problems~\cite{Gili2022EvaluatingGI}. In~\cite{Gili2022EvaluatingGI}, different sample-based metrics were introduced reflecting the generalization performances of models from a validity and a quality perspective. The latter is based on the definition of a task-specific objective or cost. 
\par
In this work, we apply TN-based generative models introduced in~\cite{tngenerative} on molecules using their SELFIES representation. There is evidence that SELFIES is a robust representation for molecules~\cite{Krenn2020}. When using SELFIES as inputs of generative models in~\cite{Krenn2020}, the validity rates were close to $100\%$ when generating new molecules. We perform training of different TN-based generative models on a subset of $4989$ molecules from the QM9 dataset~\cite{qm95k, Ramakrishnan2014} and an in-house small dataset of $516$ validated antioxidants from TotalEnergies. Our contributions can be summarized as follows:
\begin{itemize}
    \item We compare their generation quality during training using the Frechet Chemnet Distance (FCD) as a score~\cite{FCD}. The FCD has many advantages such as detecting whether generated molecules are diverse and measuring whether they have similar chemical and biological properties as molecules of interest. We compare them also to a usual GAN model as a classical model, following the example of~\cite{Gili2022EvaluatingGI}.
    \item However, differing from~\cite{Gili2022EvaluatingGI}, we compared the quality of the generations based on a metric for multiple objectives instead of a single one. We do so using the hypervolume of $3$ molecular metrics of interest per task, which is very common in multiobjective optimization~\cite{perfmulti}. We motivate this choice later in the text. On the first dataset, the GAN outperforms the TNs in terms of FCD, but not on the TotalEnergies dataset. When computing the hypervolume of $3$ molecular metrics of interest per task, the TN-based models outperform the GAN.
    \item Finally, we combine the samples from different models and study the effect of such a combination on the sample-based metrics. We find out that such a combination is beneficial for generative tasks through the small datasets used in this study. 
\end{itemize}
The structure of the paper is as follows. Section~\ref{data} describes the datasets, and how the molecular properties of interest are computed. Section~\ref{gantn} gives the necessary background on the selected models used in this work. Section~\ref{comparison} presents our results with the sampling-based metrics.  We conclude this work with a discussion in Section~\ref{discussion}.

\section{Datasets and molecular objectives}
\label{data}
In this work, we employ a subset of $4\,989$ molecules from~\cite{qm95k}, composed of all molecules with up to nine heavy atoms (C, O, N, F) from the set of $134\,000$ stable small molecules known as the QM9 dataset~\cite{Ramakrishnan2014}, and a small dataset of $516$ antioxidants from TotalEnergies validated by in-house experts. Antioxidants are compounds that inhibit oxidation, critical in a variety of industrial applications and products such as lubricants, high-efficiency gasoline, and more durable polymers.
\par
The first dataset is generally used as a small benchmark dataset for molecular generation tasks while the other reflects the small sizes of industrial molecular datasets. Despite the availability of many large molecular datasets online, 
we focus on two small ones here and aim to test the generalizability of quantum generative models trained on a few training samples, which are (i) close to the industrial scenario (few but valuable samples) and (ii) supported theoretically in previous works, e.g.,~\cite{Caro2022}.
Hence, small datasets can be an interesting case of potential practical application for quantum models once current limitations in training large and deep circuits are overcome. In our case, when considering SELFIES representations~\cite{Krenn2020}, the respective alphabets per dataset contain $33$ and $17$ letters, and SELFIES strings have up to $21$ and $58$ letters. When taking the logarithm base $2$ of the alphabet count, we would require $126$ and $290$ qubits for this problem. 
\par
The work of~\cite{Gili2022EvaluatingGI} proposed an approach for evaluating and comparing the generalization capabilities of models, whether they are classical, quantum-inspired, hybrid quantum-classical, or fully quantum. Different sample-based generalization metrics were defined and applied to compare an MPS model and a GAN with application to a combinatorial optimization problem. Three metrics were designed for evaluating generalization based on the validity of samples, and two for evaluating the quality of the models when interested in a specific task-dependent cost or metric. In our work, as we employ SELFIES, validity-based sampling is already given for molecules due to the robustness of the representation of molecules. Indeed, new samples from models are always valid, especially when trained on small datasets. However, it is possible to specify validity differently. For instance, we can impose conditions on molecular properties as we demonstrate later.
\par
To determine the performance of the generative models, we aim to quantify the optimality of those task-dependent metrics simultaneously instead of a linear scalarization thereof. When several conflicting metrics are subject to maximization, we typically take the notion of Pareto optimality to compare different models, where we say a performance value/point $a\in\mathbb{R}^m$ (assume $m$ cost metrics in general) is dominated by $b\in\mathbb{R}^m$ (denoted by $b\prec a$) if and only if $a$ is inferior to $b$ on all metrics. In multi-objective decision analysis, the notion of Pareto optimality can be measured by the widely-applied hypervolume indicator~\cite{ZitzlerTLFF03,ZitzlerT98,perfmulti}, which is defined as the Lebesgue measure (or volume) of the compact set dominated by a point set $S\subset \mathbb{R}^m$ (in our case, the performance values of samples drawn from a generative model) and cut from above by a reference point $r\in \mathbb{R}^m$:
$$
 \operatorname{HV}(S;r) = \lambda_m\left(\left\{p\in\mathbb{R}^m\colon \exists s\in S(s\prec p \wedge p\prec r)\right\}\right),
$$
where $\lambda_m$ denotes the $m$-dimensional Lebesgue measure on $\mathbb{R}^m$.
A higher hypervolume implies higher solution quality w.r.t. the simultaneous optimization of the objectives. There is also the possibility to compute the volume contribution given by a set of points to provide insights about such solutions \cite{Bringmann2010, Fonseca}.
\par 
Considering the $4\,989$ molecules from~\cite{qm95k}, we can compute a number of qualities such as the drug likeliness (QED), synthetizability (SA), and solubility (logP) scores of the samples generated by the model, from the RDKit package~\cite{rdkit2022}, similarly to~\cite{molgan}. For the small dataset of antioxidants, bond dissociation energy (BDE), ionization potential (IP), and synthetizability (SA) are of main interest. 
Lower BDE, higher IP, and low SA scores are preferred for this use case~\cite{bdems,bder2}. Unfortunately, the estimation of BDE and IP is costly, which is why we rely on surrogate deep learning models, such as Alfabet and AimNet-NSE~\cite{aimnse, bde1, bde2}, that can predict these properties. When computing these properties, we define two sets of criteria for considering samples. The first set of criteria, denoted (i), imposes that the generated molecules must have at least one O-H bond, and present strictly positive BDE and IP scores from the models. The second set, denoted (ii), is a recommendation from molecular experts, requiring additionally that a molecule present an IP higher than $182$ kcal/mol and a BDE lower than $85$ kcal/mol.

\section{Description of the models used}
\label{gantn}

In this section, we describe the generative models used in this study.
We first present the GAN framework, and briefly mention how it was applied to molecular generation, especially with SELFIES. Then we enumerate the different tensor-network-based models. Finally, we describe how we performed the hyperparameter search to perform the comparison of these different models for the next section.

\subsection{GAN}
GANs~\cite{gan} are very popular generative machine learning algorithms that have been quite successfully applied in image generation. A generator $G_{\vec{\theta}}$ and a discriminator $D_{\vec{\phi}}$ are trained in an alternating fashion, with their parameters being updated using Stochastic Gradient Descent algorithms such as Adam~\cite{adam}. 
$G$ learns to map samples $z$ from a prior distribution $q$ (in practice the normal one) to good outputs that can \guillet{fool} the discriminator $D$, distinguishing real data $x$ from fake. 
\begin{align}
C_{\text{GAN}} = \min_{\vec{\theta}} \max_{\vec{\phi}}  [E_{x \sim P_{\text{Train}}}(x)[\log D_{\vec{\phi}}(x)] \nonumber\\ 
+ E_{z \sim q(z)}[\log (1 - D_{\vec{\phi}}(G_{\vec{\theta}}(z)))]]
\end{align}
However, the training can be unstable due to the two models competing with each other. Most often, in practice, the generator and discriminator are designed such that their architectures do not differ drastically. Otherwise, this enables unhealthy competition between them, with one \guillet{superseding} the other in the generation game. In the context of molecular generation, different GAN approaches have been employed such as MolGan~\cite{molgan} to generate small molecular graphs or even for 3D representations~\cite{9973452}, combined with reinforcement learning to optimize metrics such as validity, novelty, or desired chemical properties obtained with SMILES and RdKit~\cite{rdkit2022}. With SELFIES, it was demonstrated that the generation quality of generative models was improved, especially in terms of validity and novelty.

\subsection{Tensor Networks for unsupervised learning}

Tensor Networks are powerful representations of high-dimensional tensors, which can be used for modeling discrete multivariate probability distributions. Given $N$ discrete random variables $\{X_i\}_i$ taking values in $\{1,\ldots,d\}$, 
we use a tensor $T\in V^{\otimes N}, V=[0, 1]^d$ (after normalization) to store the probability distribution $P(X_1,\ldots,X_N)$.
The goal during training is then to learn the underlying probability distribution $P(X_1,\ldots,X_N)$ for a dataset ${\{x_t\}^{D}_{t = 1}}$, where $x_t$ is a realization of the random variables $\{X_i\}_i$
by minimizing a log-likelihood cost function over a set of parameters $\theta$ for the TN: $- \frac{1}{D} \sum_{t} \log(p_{\theta}(x_t))$. 
\par 
Different TN models have been introduced in~\cite{tngenerative}. As shown in~\cite{tngenerative}, they exhibit tractable likelihoods and admit efficient learning algorithms. Following the notation of~\cite{tngenerative}, we present the three different TN representations of the entries of $T$ using real or complex tensors:
\begin{enumerate}
\item \textbf{Positive tensor-train/matrix product state (MPS)}: 
$$T_{X_1,\ldots, X_N} = \sum_{\{\alpha_i=1\}}^{r} A^{\alpha_1}_{1,X_1}A^{\alpha_1, \alpha_2}_{2,X_2}\cdots A^{\alpha_{N-2},\alpha_{N-1}}_{N-1,X_{N-1}}A^{\alpha_{N-1}}_{N,X_N}\ ,$$
where $A_1$ and $A_N$ are $d\times r$ non-negative valued matrices, and $A_i$ are order-$3$ tensors of dimension $d\times r\times r$, with elements in $\mathbb{R}_{\geq 0}$. The indices $\alpha_i$ of these constituent tensors are contracted (summed over) to construct $T$.
\item \textbf{Born machine (BM)}:
$$T_{X_1,\ldots, X_N} = \left|\sum_{\{\alpha_i=1\}}^{r} A^{\alpha_1}_{1,X_1}A^{\alpha_1 ,\alpha_2}_{2,X_2}\cdots A^{\alpha_{N-2},\alpha_{N-1}}_{N-1,X_{N-1}}A^{\alpha_{N-1}}_{N,X_N}\right|^2,$$
with elements of the constituent tensors $A_i$ in $\mathbb{F}\in\{\mathbb{R},\mathbb{C}\}$. 
\item \textbf{Locally purified state (LPS)}:
\begin{align}
T_{X_1,\ldots, X_N} = \sum_{\{\alpha_i,\alpha'_i=1\}}^{r} \sum_{\{\beta_i=1\}}^\mu   \nonumber\\
A^{\beta_1,\alpha_1}_{1,X_1}\overline{A^{\beta_1,\alpha'_1}_{1,X_1}} A^{\beta_2, \alpha_1 ,\alpha_2}_{2,X_2} \overline{A^{\beta_2, \alpha'_1 ,\alpha'_2}_{2,X_2}}\cdots A^{\beta_{N},\alpha_{N-1}}_{N,X_N}\overline{A^{\beta_{N},\alpha'_{N-1}}_{N,X_N}}
\end{align}
with elements of the constituent tensors $A_i$ in $\mathbb{F}\in\{\mathbb{R},\mathbb{C}\}$. $A_1$ and $A_N$ are order-$3$ tensors of dimension $d \times \mu \times r$ and $A_i$ are order-$4$ tensors of dimension $d \times \mu \times r \times r$. The indices $\alpha_i$ run from 1 to $r$, the indices $\beta_i$ run from 1 to $\mu$.
\end{enumerate}
The previous models can have corresponding quantum circuit representations as demonstrated in~\cite{tngenerative}. Hence, these quantum-inspired models can be tried before applying a full quantum model once current challenges in training large and deep circuits are overcome as was done for combinatorial optimization problems~\cite{Gili2022EvaluatingGI}. 

\subsection{Hyperparameter selection for model comparison}

For comparing a GAN model where we use neural networks similar to~\cite{Krenn2020} and TNs, we first performed a random search of hyperparameters to find a good set of values for each model. At each epoch, we sample $10\,000$ SELFIES, then convert them to SMILES, and compute the FCD w.r.t. the SMILES representation composing the dataset. We then keep the samples from the epoch in terms of lower FCD. We use a similar GAN procedure used in~\cite{Krenn2020}, where the generator $G$ and the discriminator $D$ are defined as multi-layer perceptrons. While the latter are defined with one hidden layer in~\cite{Krenn2020}, we use up to $3$ hidden layers in our case. The architectures of $G$ and $D$ share the same number of hidden layers and units. We used Adam as optimizer~\cite{adam} during training, and apply dropout~\cite{JMLR:v15:srivastava14a} when training the discriminator. The hyperparameters considered for random search are the learning rate values for Adam (uniformly sampled on a log-scale between $-7$ and $-4$), the number of hidden units (randomly sampled between $300$ and $3000$), the dimension of the prior (between $50$ and $300$), and the dropout rate (between $0$ and $0.8$). We sample $200$ sets of hyperparameter values per number of layers when performing random search. For TN, we use the code from~\cite{tngenerative}, and try different $3$ values of $r$, that is $2, 3, 5$. All TN models are trained under a budget of $200$ epochs and $1000$ for the GAN (training performances would not improve by increasing the budget in our case). More hyperparameters specifications and generative models can be tried in future works, but the comparison will be made in a similar way as we present in the next section.

\section{Comparison results}
\label{comparison}

Having previously defined the models to be compared, the datasets, and the objectives per task and dataset, we present our results in this section. First, we study their learning capabilities w.r.t. the molecular datasets. Then, for the antioxidants tasks, we apply the metrics from~\cite{Gili2022EvaluatingGI} reflecting the validity-based generalization capabilities of the models. Finally, on all tasks, we compare models by computing the hypervolume over $3$ properties of interest. 

\subsection{Learning capabilities}
For each model trained on each dataset, we investigate their learning capabilities by computing the lowest Frechet Chemnet Distance~\cite{FCD} of the generated data. Table~\ref{quantum-table-mols} present our results.
On the QM9 subset, the GAN performed best in training performances. However, on the antioxidants dataset, we witness the GAN being outperformed by TNs. In particular, we achieved a lower FCD with the RealBorn model, and the RealLPS in median when considering $5$ training runs. This indicates an interesting setting for considering quantum models. 

\begin{table}[ht]
\caption{Minimal, median, and standard deviation of the FCD obtained over $5$ runs and $200$ epochs.}
\label{quantum-table-mols}
\vskip 0.15in
\begin{center}
\begin{small}
\begin{sc}
\begin{tabular}{llllr}
\toprule
Model/QM9 & Min FCD & Median  & Std\\
\midrule
PositiveMPS  & 7.869 & 9.887 & 2.260 \\
RealBorn  &  6.175 &  9.022 & 1.567 \\
RealLPS  & 6.522 & 7.795 & 1.604 \\
ComplexBorn  & 7.081 & 8.574 & 1.078 \\
ComplexLPS  & 7.223 & 7.897 & 1.146 \\
GAN & \textBF{5.301} & \textBF{5.300} & 0.486 \\
\midrule
Model/Antioxidants & & & \\
\midrule

PositiveMPS & 35.935  & 37.923 & 2.841 \\
RealBorn  & \textBF{35.799} & 37.512 & 0.921 \\
RealLPS  & 36.127 & \textBF{36.397} & 0.585 \\
ComplexBorn  & 35.803 & 36.408 & 0.167 \\
ComplexLPS & 36.217 & 36.616 & 0.316 \\
GAN & 39.696 & 40.613 & 0.068\\

\bottomrule

\end{tabular}
\end{sc}
\end{small}
\end{center}
\vskip -0.1in
\end{table}

\subsection{Validity-based generalization for antioxidants}

\begin{figure*}[ht]
\vskip 0.2in
\centering
\subfloat[]{\includegraphics[width=0.33\textwidth]{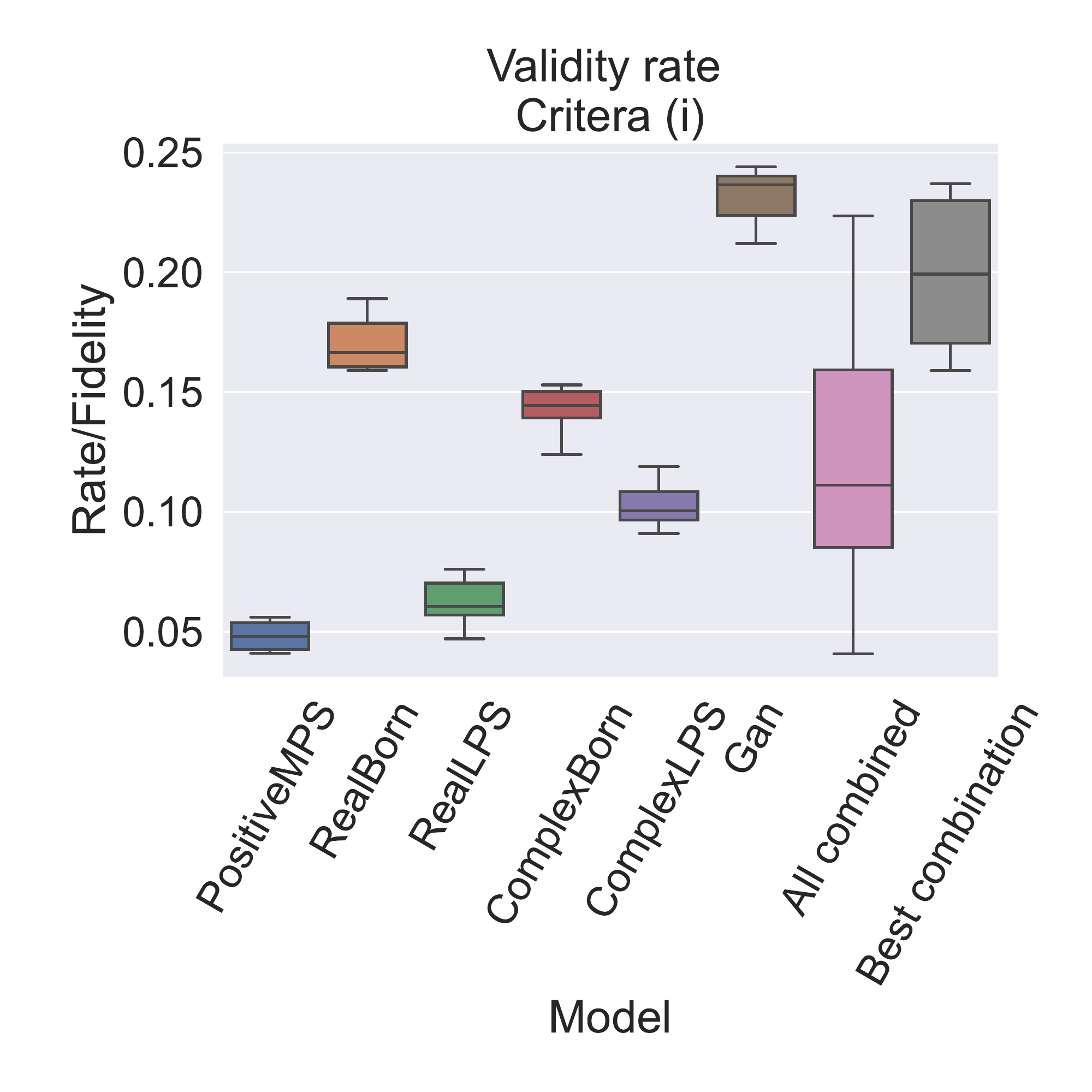}}
\subfloat[]{\includegraphics[width=0.33\textwidth]{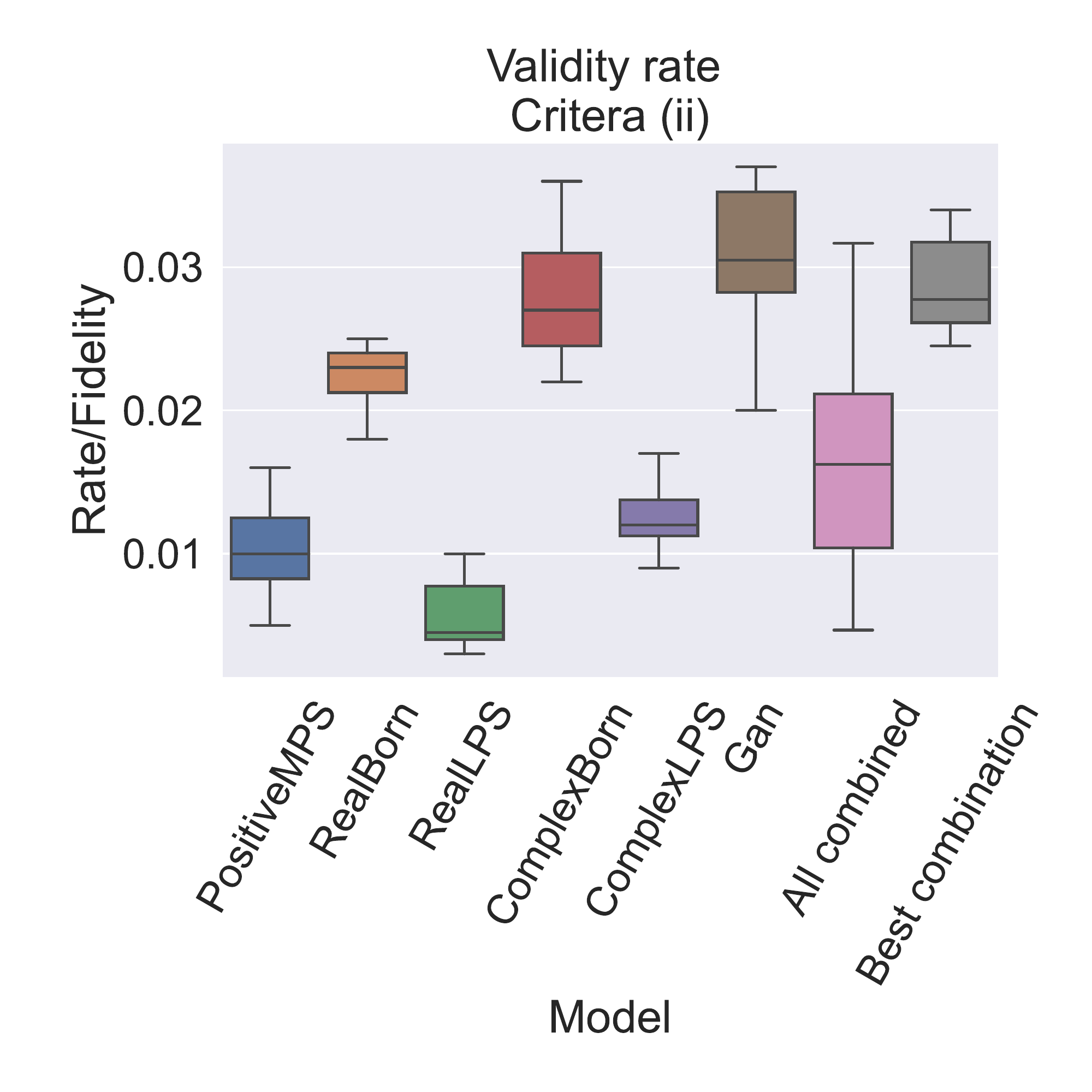}}
\caption{ \small Validity rate for the set of criteria (i) (left) and (ii) (right).}
\label{fidelitymols}
\vskip -0.2in
\end{figure*}

Here, we present the results from one of the metrics introduced in~\cite{Gili2022EvaluatingGI} applied to the antioxidants dataset, i.e. fidelity. The latter is defined as the ratio of the number of new and valid (w.r.t. to the sets of criteria) samples over new (and possibly invalid) samples w.r.t. the antioxidant dataset. Note that in~\cite{Gili2022EvaluatingGI}, another metric called rate, is defined as the ratio of the number of new and valid samples over the number of samples drawn. In our case, the fidelity is equal to the rate as all samples drawn differ from the dataset. Fig.~\ref{fidelitymols} presents the fidelity from sampling $1\,000$ molecules $10$ times each model, and also from their combination, for criteria (i) and (ii). We do so by dividing in $10$ folds the best generations by a model in terms of FCD. Firstly, we witness that GAN outperforms TNs for criteria (i), realizing a median fidelity of $23.65\%$, and similarly on criteria (ii) with $3.05\%$. However, when combining models, for both sets of criteria, such a ratio is averaged between less and more performing models. Hence, the combination can be beneficial in balancing the efficiency of the models in distinguishing between unseen valid and invalid candidates. 

\subsection{Hypervolume of molecular metrics}

\begin{figure*}[!ht]
\centering
\subfloat[]{\includegraphics[width=0.33\textwidth]{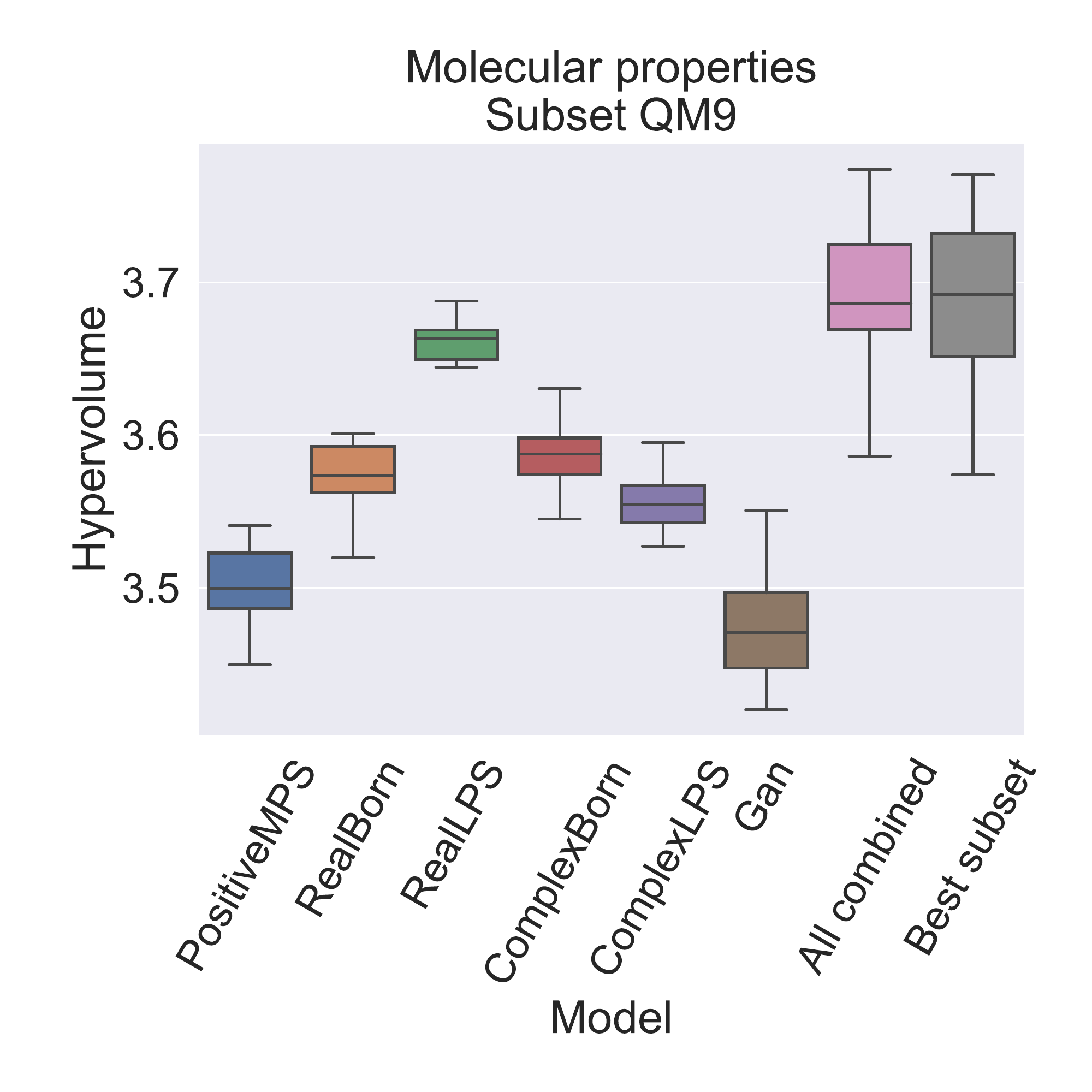}}
\subfloat[]{\includegraphics[width=0.33\textwidth]{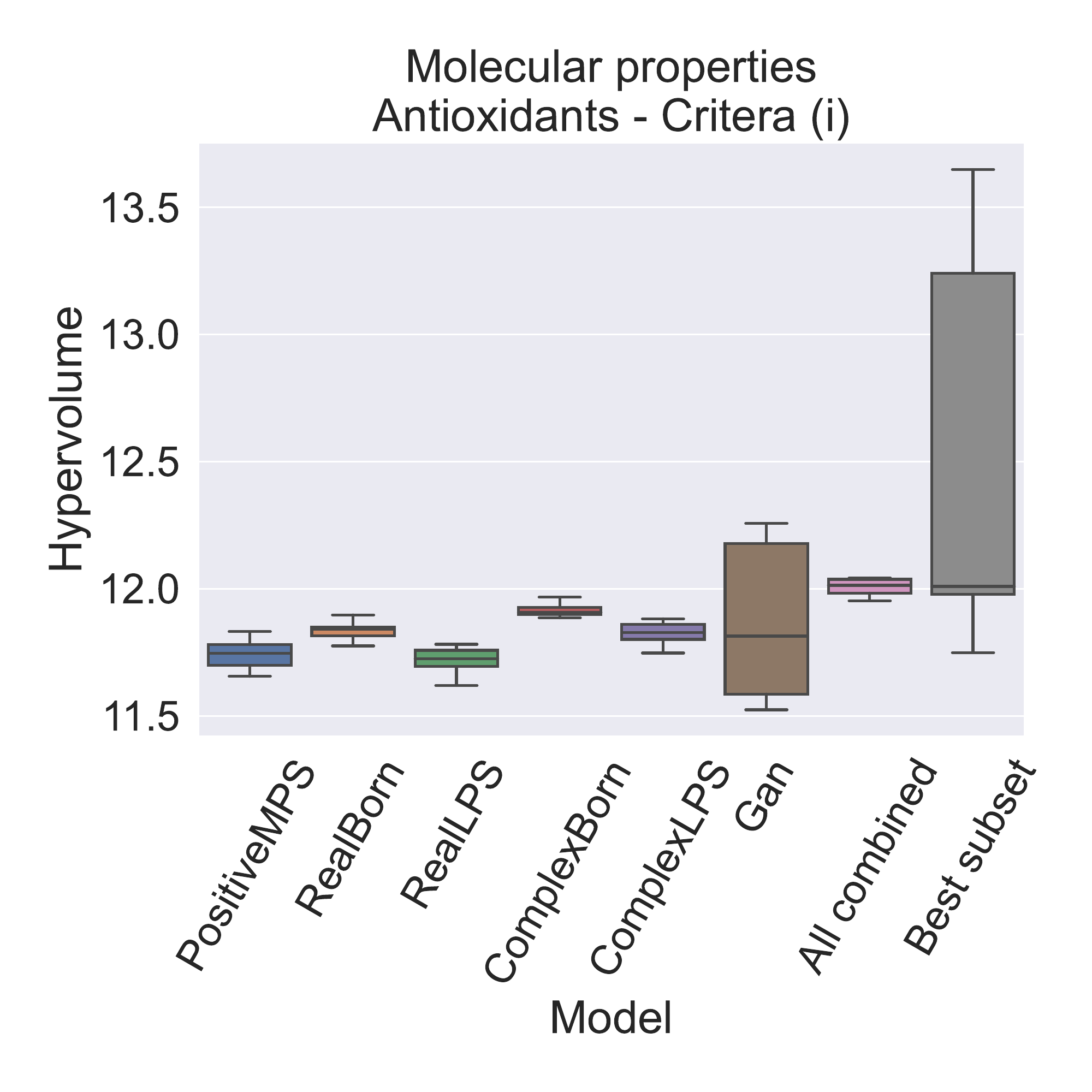}}
\subfloat[]{\includegraphics[width=0.33\textwidth]{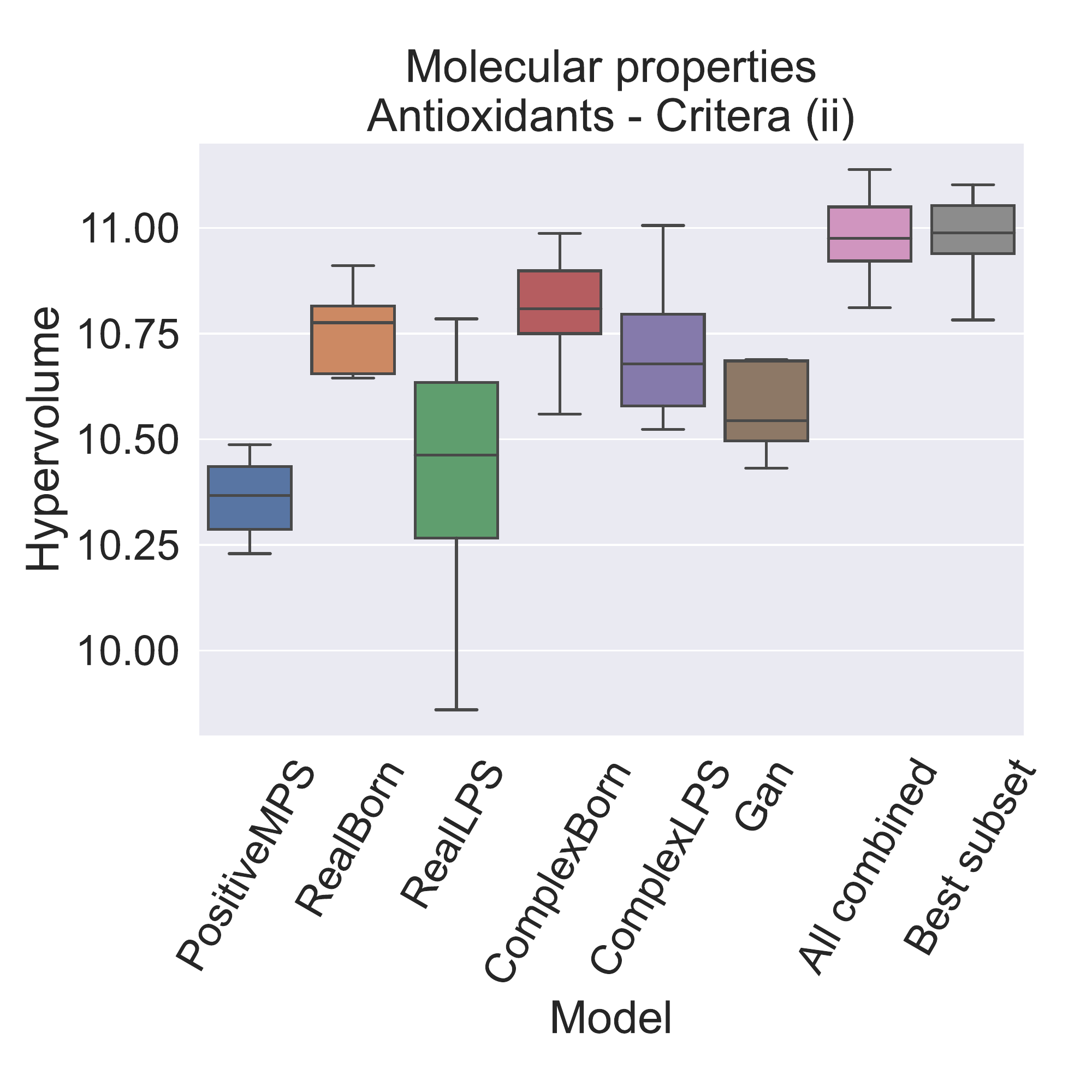}}
\caption{ \small Hypervolumes (log-scale) of dividing in $10$ folds the best generations by a model in terms of FCD for $3$ computed metrics for the QM95k dataset (left) and the antioxidants (middle for (i)- right for (ii)). We take as a reference point ($0, -7, 1$) for drug likeliness, solubility, and synthetizability, to be able to compute the hypervolume with the obtained samples.
Similarly, for the antioxidants tasks, we take ($140, 0, 10$) for (i) and ($85, 182, 10$) for (ii), for bond dissociation energy, ionization potential, and synthetizability.}
\label{hvolres}
\end{figure*}

Here, we present the multiobjective results from sampling $1000$ molecules $10$ times each model, and also from their combination (again dividing in $10$ folds the best generations by a model in terms of FCD). Fig.~\ref{hvolres} presents our results where we compute the hypervolume (presented in log scale) over the $3$ metrics of interest computed for each sample. Concerning the reference points chosen, we select them based on the range of property values we obtained across all generations. A higher hypervolume value means better coverage in terms of optimizing these $3$ metrics simultaneously. We witness that:
\begin{enumerate}
    \item For the subset of QM9 molecules, GAN is outperformed by the TNs, RealLPS being the best-performing model. RealLPS achieves a median of $3.663$, while for GAN it is lower ($3.471$). If we correlate to our results from Table~\ref{quantum-table-mols}, as GAN has a higher learning capability for this task, its generation quality is impacted. RealLPS is the second lowest median FCD though, but its generation quality is still competitive against the other TNs. Hence, one should balance learning and generalizing in the quality of sampling.
    \item For criteria (i) and (ii) on the antioxidant dataset, the ComplexBorn model performs better in terms of median hypervolume (respectively, $11.907$ and $10.809$). Yet, for criteria (i), the generations produced by the GAN model can achieve higher values, albeit lower median ($11.815$). If we correlate again to our results from Table~\ref{quantum-table-mols}, ComplexBorn is the second lowest median FCD after RealLPS, which achieves lower hypervolume. This again points out to balancing learning and generalizing in the quality of sampling.
    \item As for the combination of all models, we witness again a balancing effect and even an improvement in the quality of candidates vis-a-vis of the $3$ molecular objective per task. Respectively per task displayed in Fig.~\ref{hvolres}, the median hypervolume is increased, in absolute difference from the best model individually, by $0.02, 0.11$, and $0.17$. 
    \item Finally, we show also the hypervolumes when combining a subset of the models. We select the best combination in terms of hypervolume by trying all possible subsets. The median hypervolume is slightly increased compared to combining all models for the QM9 and criteria (ii) tasks. For the QM9 task, $4$ models were selected leaving the $2$ least performing ones in terms of median hypervolume. They correspond to the GAN and the MPS models. For criteria (ii), RealLPS was left out as the second least performing. For criteria (i), although the median is slightly decreased by removing the ComplexLPS model (which is not in the least-performing models in terms of hypervolume), we witness we can obtain higher hypervolumes.
\end{enumerate}

In conclusion, we find that GAN can be outperformed in general by a TN model in terms of sample-objective quality. But balancing learning and general quality of sampling is crucial, and combining the samples of multiple generative models is an interesting solution to achieve so.

\section{Discussion}
\label{discussion}

In this work, we applied different quantum-inspired models introduced in~\cite{tngenerative} and compare them with a conventional classical GAN model, on two small molecular datasets, one used as a small benchmark while the other is being provided by industry. We use SELFIES representation when training the generative models. Our comparison is done using different sample-based metrics similarly to~\cite{Gili2022EvaluatingGI}, reflecting their learning performances, and multiobjective performances using $3$ molecular metrics of interest per task. We find that GAN can be outperformed in general by a TN model similar to~\cite{Gili2022EvaluatingGI}. We also used an indicator from multiobjective optimization, the hypervolume, to compare the quality of samples from different models vis-a-vis different molecular metrics. The results highlight the importance of balancing learning and the general quality of sampling, especially when we combine different models. Unifying classical, quantum-inspired, fully quantum, or/and even other more specialized hardware become an interesting direction for generative modeling. Future works can use similar comparison approaches of generative models, considering hybrid quantum-classical approaches, incrementally increasing the quantum part toward fully quantum models that can provide an advantage in sampling. Finally, it is also possible to rely on optimizing the metrics when training generative models, using for instance reinforcement learning as it was done with MolGan.

\acknowledgements 

CM and VD acknowledge support from TotalEnergies. CM and MAP would also like to thank Huanyi Qin from SUNY Binghamton who let us use his code for computing the BDE and IP. This work was supported by the Dutch Research Council (NWO/OCW), as part of the Quantum Software Consortium programme (project number 024.003.037). This research is also supported by the project NEASQC funded from the European Union’s Horizon 2020 research and innovation programme (grant agreement No 951821).

\appendix

\bibliographystyle{icml2022}
\bibliography{references}

\end{document}